\documentclass[epjST]{svjour}
\usepackage{graphics}
\usepackage{cite}
\usepackage{amsmath}
\usepackage{amssymb}
\usepackage{latexsym}
\usepackage{amsfonts}
\usepackage{epsfig}
\usepackage{psfrag}
\usepackage{graphicx}
\usepackage{color}

\newcommand{\nn}{\nonumber\\}

\newcommand{\bea}{\begin{eqnarray}}
\newcommand{\ea}{\end{eqnarray}}
\newcommand{\eea}{\end{eqnarray}}
\newcommand{\ord}{\,{\cal O}}
\newcommand{\li}{\,\widehat{\cal L}}
\newcommand{\tr}{\,{\rm Tr}}

\begin{document}
\title{Classical spin simulations with a quantum two-spin correction}
\author{Patrick Navez\inst{1,2} \fnmsep\thanks{\email{p.navez@hzdr.de}} \and Grigory A. Starkov\inst{1,3} \and Boris V. Fine\inst{1,4}  }
\institute{ 
Skolkovo Institute of Science and Technology, 
Skolkovo Innovation Centre, Nobel Street 3, Moscow 143026, Russia 
\and 
University of Saskatchewan, Dept of Math. and Stat. , Saskatoon, S7N 5E6, Canada
\and
Lebedev Physical Institute of the Russian Academy of Sciences, Leninsky prospect 53,
Moscow 119991, Russia
\and
Institute for Theoretical Physics, University of Heidelberg, Philosophenweg 12, 69120 Heidelberg, Germany
}
\abstract{
Classical simulations of high-temperature nuclear spin dynamics in solids are known 
to accurately predict relaxation for spin 1/2 lattices with a large number of interacting neighbors. 
Once the number of interacting neighbors becomes four or smaller, classical simulations  
lead to noticeable discrepancies. Here we attempt to improve the performance of the classical 
simulations by adding a term representing two-spin quantum correlations. The method is tested for a spin-1/2 chain. 
It exhibits good performance at shorter times, but, at longer times, 
it is hampered by a singular behavior of the resulting equations of motion.
} 
\maketitle

%
\section{Introduction}
\label{intro}

Nuclear spin-spin relaxation in solids has been a subject of active theoretical 
research\cite{Bloch-46-1,VanVleck-48,Lowe-57,Abragam-61,Tjon-66,Gade-66,Borckmans-68,Jensen-73,Parker-73,Jensen-73A,
Engelsberg-74,Engelsberg-75,Becker-76,Jensen-76,Lundin-77,Shakhmuratov-91,Lundin-92,Tang-92,Jensen-95,Sodickson-95,
Fine-97,Fine-00,Dobrovitski-06,Zhang-07,Elsayed-13-thesis,Elsayed-13,Elsayed-15} since the 
discovery of nuclear magnetic resonance (NMR) \cite{Purcell-46,Bloch-46-2}. 
Nuclear spin dynamics is usually considered in the limit of infinite temperature as 
the nuclear gyromagnetic ratios and, hence, the interaction energies are very small. Despite this simplification, 
the dynamics is still nontrivial. Making controllable first-principles predictions of 
NMR relaxation has remained an elusive goal for the practitioners in the field.

Classical spin simulations were shown to be rather accurate in predicting 
the NMR spin-spin relaxation for the lattices of quantum spins 1/2 with a 
large number of interacting neighbors\cite{Jensen-73,Jensen-76,Lundin-77,Tang-92,Elsayed-13-thesis,Elsayed-15}. 
However, as the number of the interacting neighbors decreases to 
four or smaller, classical simulations lead to noticeable discrepancies. In a broader context, similarity 
and contrast between quantum and classical spin dynamics were also investigated. On the one hand, there exists 
significant experimental and numerical evidence supported by theoretical 
arguments\cite{Borckmans-68,Engelsberg-74,Fabricius-97,Fine-03,Fine-04,Fine-05,Morgan-08,Sorte-11,Meier-12,Steinigeweg-12} that 
the asymptotic high-temperature relaxation in ergodic classical and quantum spin systems has the functional 
form either exp$(- \gamma t)$ or exp$(- \gamma t)$~cos$(\omega t + \phi)$, where $\gamma$, $\omega$ and $\phi$ are some constants. 
On the other hand, the classical spin dynamics is known to be chaotic \cite{deWijn-12,deWijn-13, Fine-14,Elsayed-15A}, while 
the dynamics of spin-1/2 lattices was conjectured in Refs.\cite{Fine-14,Elsayed-15A} to be non-chaotic in the sense of not exhibiting exponential sensitivity to small perturbations.
 
In this work, we attempt to improve classical simulations of quantum spin dynamics by introducing corrections representing purely quantum correlations between each pair of interacting spins. The correction is introduced on the basis of an expansion in terms of
the inverse of the lattice coordination number \cite{NS10,QKNS14}. This expansion was already used to describe the quantum dynamics of other lattice systems such as  
the Bose- and Fermi-Hubbard models \cite{NS10,QKNS14,NQS14,KNQS14} and the Heisenberg spin model \cite{NTZ17,NQS14}.
 An alternative approach to combining classical and quantum simulations was recently introduced  by two of us in Ref. \cite{greg}.

In section 2, we give general formulation of the NMR spin-spin relaxation problem. 
In section 3, we present the method of large coordination number expansion, from which  we derive the dynamical equations for the spin average and correlations. In section 4, 
we  present the results of applying our method to a chain of spins 1/2. Section 5 contains the conclusions.

\section{General formulation}

We consider a $D$-dimensional hypercubic lattice of spins 1/2 described by the nearest-neighbor interaction Hamiltonian
\bea
\hat H&=&\frac{1}{Z}\sum_{m,n}\left( J_{mn}^{x}
\hat{S}_{m}^x \hat{S}_{n}^x+J_{mn}^{y}
\hat{S}_{m}^y \hat{S}_{n}^y +J_{mn}^{z}
\hat{S}_{m}^z \hat{S}_{n}^z \right)
\,.
\ea
where $\hat{S}_{n}^i$ ($i=x,y,z$)  are the spin projection operators for site $n$,   $J_{mn}^{i}$  are the coupling constants for the $i$th projections of spins $m$ and $n$. 

The quantity of interest in the context of NMR is the infinite-temperature autocorrelation function defined 
as \cite{Lowe-57,Abragam-61}:
\begin{eqnarray}\label{correlation}
C(t)= \frac{\tr\left(\hat S^x(t) \hat S^x\right)}
{\tr\left(\hat S^{x\,2}\right)}\, ,  
\end{eqnarray}
where $\hat S^x \equiv \sum_m \hat S^x_m$ is the $x$-component of the total spin polarization. Its time dependence is defined in the Heisenberg representation as
\begin{equation}
 \hat S^x(t)=\exp(i\hat H t/\hbar ) \hat S^x \exp(-i\hat H t/\hbar ) 
 \label{Sx}
 \end{equation}
Function $C(t)$ is proportional to the signal of NMR free induction decay.

A previous work  \cite{Elsayed-15} obtained $C(t)$ by classical simulations for one-, two- and three-dimensional lattices. 
For two-  and three-dimensional lattices the classical results exhibited reasonable agreement either with the direct quantum 
calculation or with experiment.  However, for one-dimensional chain, the agreement was not very good. Here we aim at achieving 
an improvement in the latter case by 
taking into account the next order quantum corrections describing the two-spin correlations. 

In order to derive the quantum correction, we perform an expansion in terms of the inverse powers of 
the lattice coordination number $Z=2D$.  When applied up to the  order $1/Z^2$, this method showed a reasonable convergence 
in the case of  the Bose-Hubbard lattice model in one and two dimensions\cite{KNQS14}.
It has also been used successfully for the quantum Ising model \cite{NQS14,NTZ17}.
In the present work, the resulting approximation is to be tested for the worst possible case, namely, for a one-dimensional chain (Z=2).

\section{Large coordination number expansion}

{\subsection{Overview of the method}} 

For $Z\gg1$, 
the model dynamics is  to be described using the method developed 
in \cite{NS10,QKNS14}, which we now introduce.

The time evolution of the density matrix $\hat\rho$ of the whole lattice is 
given by the von Neumann-Liouville equation 
$
i\hbar \partial_t\hat\rho 
=
\left[\hat H,\hat\rho\right] 
$.
This density matrix is usually too complex to be analyzed. Instead, the 
set of reduced density matrices is introduced, 
$\hat\rho_{\cal S}=\tr_{\not {\cal S}}\hat \rho$, which results from tracing out the Hilbert spaces of all 
sites except a few: ${\cal S}=\{n_1,n_2, \dots, n_i\}$. If we keep only one site $n$, 
then the reduced density matrix is a linear operator $\hat \rho_n$ 
acting on the Hilbert space 
of one lattice site $n$;  if we keep two sites $n,m$, then $\hat\rho_{n,m}$ acts on the Hilbert space of two sites etc.
%
The decomposition 
%
$\hat\rho_{m,n}
=
\hat\rho_{m,n}^{\rm corr}+\hat\rho_{m}\hat\rho_{n}
\,,
$
and 
$\hat\rho_{m,n,r}=\hat\rho_{m,n,r}^{\rm corr}+
\hat\rho_{m,n}^{\rm corr}\hat\rho_{r}+
\hat\rho_{m,r}^{\rm corr}\hat\rho_{n}+
\hat\rho_{n,r}^{\rm corr}\hat\rho_{m}+
\hat\rho_{m}\hat\rho_{n}\hat\rho_{r}$, 
etc. allows us to derive an exact hierarchy of interlinked equations for 
these operators \cite{NS10,QKNS14}. This hierarchy is the counterpart of the Bogoliubov-Born-Green-Kirkwood-Yvon (BBGKY) chain but with correlations 
between sites and not between particles.

%
In order to treat the time dynamics, it was demonstrated quite generally in \cite{NS10,QKNS14} that, if the initial state of 
any quantum lattice system is separable without initial correlations between sites, then the 
correlations at a later time satisfy -- at least for a finite period of time -- 
the following hierarchy scaling
$\hat \rho^{\rm corr}_{\cal S} \sim 1/Z^{|{\cal{S}}|-1}$, which means that the higher-order 
correlations are suppressed as an inverse power of the coordination number $Z$.  More explicitly, 
\bea \label{rho}
\label{hierarchy}
\hat\rho_{n} = \ord\left(Z^0\right)
,\,
\hat\rho^{\rm corr}_{m,n} = \ord\left(1/Z\right)
,\,
\hat\rho^{\rm corr}_{m,n,r} = \ord\left(1/Z^2\right)
,
\ea
and so on.
Using the spin representation, this hierarchy can be rewritten as 
\begin{eqnarray}\label{dS}
S^{\alpha}_{n}
=\langle \hat S^{\alpha}_{n} \rangle &=& \ord \left(Z^0\right)
,\ \ 
M_{m,n}^{\alpha\beta}=\langle \delta \hat S^{\alpha}_{m} \delta \hat S^{\beta}_{n} \rangle 
=\ord \left(1/Z\right)
,
\nonumber \\
\langle \delta \hat S^{\alpha}_{m} \delta \hat S^{\beta}_{n} \delta \hat S^{\gamma}_{r} \rangle 
&=& \ord \left(1/Z^{2}\right), \ \ \dots  \quad \quad  \alpha, \beta, \gamma=x,y,z~,
\end{eqnarray}
where, for an operator $\hat A$,  $\delta \hat A = \hat A -\langle \hat A \rangle$ and 
$\langle \hat A \rangle=\tr(\hat A \hat \rho)$. 

Many methods of quantum field theory use similar expansion techniques such as 
the $1/{\cal N}$ expansion, where ${\cal N}$ is the number of field components, or the
$1/S$ expansion, where $S$ is the quantum spin.  
The open question is whether this kind of expansion converges accurately towards 
the exact solution. In our case, we will expand only up to the 
first order and test the resulting approximation.

In \cite{NS10,QKNS14}, an exact set of hierarchy equations has been 
derived for these density matrices. Up to the first order, these are written as:
\begin{eqnarray}
\label{one-site-approx}
i\partial_t\hat\rho_{n}
&=&
\frac{1}{Z}
\sum_{m\neq n}\tr_{m}\left\{
\li^S_{n,m}\left(
\hat\rho_n \hat\rho_m + \hat\rho^{\rm corr}_{n,m}\right) \right\}
\\
\label{two-sites-approx}
i \partial_t \hat\rho^{\rm corr}_{n,m}
&=&
\frac1Z\li_{n,m}(\hat\rho_m\hat\rho_n+\hat\rho^{\rm corr}_{n,m})
-\frac{\hat\rho_{n}}{Z}
\tr_{n}
\left\{\li^S_{n,m}(\hat\rho_m\hat\rho_n+\hat\rho^{\rm corr}_{m,n})\right\}
\nn
&&+
\frac1Z
\sum_{r\not=n,m} 
\tr_{r}
\left\{
\li^S_{n,r}
(\hat\rho^{\rm corr}_{n,m}\hat\rho_{r}+
\hat\rho^{\rm corr}_{m,r}\hat\rho_{n})
\right\}
+(n\leftrightarrow m)
+\ord(1/Z^2)
\,
\end{eqnarray}
where we define the Liouville operators in terms of the commutators:
\begin{eqnarray}
\li_{n,m} \hat A \equiv \left[
\sum_{\alpha=x,y,z} J^{\alpha}_{m,n}
\hat{S}_{m}^\alpha \hat{S}_{n}^\alpha, \hat A \right]  \quad ,
\end{eqnarray}
and $\li^S_{m, n}=\li_{m, n}+\li_{n, m}$. Note that the 
trace in the first line of  Eq.(\ref{two-sites-approx}) is carried out on the Hilbert space associated to 
the site $n$ without a summation over the index $n$.
This set forms the basis of the $1/Z$-expansion up to the first order.
The term $\hat\rho^{\rm corr}_{n,m}$ in Eq.(\ref{two-sites-approx}) describes quantum pair correlations beyond those determined by single-spin density matrices
$\hat\rho_m$ \cite{NQS14}. 

\subsection{General set of equations including the two-spin quantum correlations}

The averages defined in Eqs.(\ref{dS}) are used into the Eqs(\ref{one-site-approx}-\ref{two-sites-approx}) in order to 
determine their dynamics evolution.
Assuming $J_{mn}^{\alpha}=J_{nm}^{\alpha}$, we arrive at the following set of expectation values for the single spin and pair operators: 
\begin{eqnarray}\label{S}
 \partial_t S_{m}^\alpha
&=& 2
\sum_{r}\displaystyle \frac{J_{mr}^{\gamma}}{Z}\epsilon^{\alpha\gamma\beta}
(S_{m}^\beta  S_{r}^\gamma + M^{\beta\gamma}_{mr}) ,
\end{eqnarray}
\begin{eqnarray}\label{M}
 \partial_t M_{mn}^{\alpha\beta} 
&=& 
2\sum_{r\not= m,n}[\displaystyle \frac{J_{mr}^\gamma}{Z}
\epsilon^{\alpha \gamma \delta}
(M_{mn}^{\delta \beta}  S_{r}^\gamma + M^{\beta \gamma}_{nr}S_m^\delta)+
\displaystyle \frac{J_{nr}^\gamma}{Z}
\epsilon^{\beta\gamma\delta}
(M_{mn}^{\alpha\delta}  S_{r}^\gamma + M^{\alpha \gamma}_{mr}S_n^\delta ) ]
\nonumber \\
&+&
2\frac{J_{mn}^\gamma}{Z}\{
\epsilon^{\alpha \gamma\delta }[(\delta^{\beta \delta }/4-S_n^{\beta} S_n^\delta)S_m^\gamma-M_{nm}^{\delta \gamma} S_{n}^\beta]
\nonumber \\
&+&
\epsilon^{\beta \gamma \delta }[(\delta^{\alpha \gamma}/4-S_m^\alpha S_m^\gamma)S_n^\delta - M_{nm}^{\delta \gamma} S_{m}^\alpha]\},
\end{eqnarray}
where $\epsilon^{\alpha \gamma \delta}$ is the anti-symmetric Levi-Civita tensor.

Equations (\ref{S}) are identical to the equation of motion for the classical spins, 
when the 
 correlation
terms $M^{\beta\gamma}_{mr}$ are removed. Together, the self-consistent system (\ref{S},\ref{M}) 
describes the dynamics of classical spin and the quantum correlations between distant spin pairs.  

We note that Eqs.(\ref{S},\ref{M}) preserve the conservation laws such as the total energy average, the total spin projection average on the z-axis 
and  the average of the total spin squared (if conserved by ${\hat H}$). 
This is achieved by keeping in Eq.(\ref{M}) the terms that do not have summation over index $r$ and hence scale as $1/Z^2$.
We also show in appendix in \ref{A2} that 
Eqs.(\ref{S},\ref{M}) lead to the exact result for the case of two spins, which is an indicator 
of the quantitative promise of the simulation scheme\cite{Fine-97}.

Equation (\ref{S}) without correlation terms $M^{\beta\gamma}_{mr}$ preserves the length of the individual spin, follows classical Hamiltonian dynamics 
and, therefore,  
does not display any dynamical instability. Once the correlation terms are included, the individual 
spin lengths are no longer conserved, the dynamics loses the Hamiltonian character, and, as a result, dynamical instabilities  eventually set in, accompanied by negative eigenvalues of the 
reduced density matrices for individual spins. 
For large $Z$, we expect that the correlation terms $M^{\beta\gamma}_{mr}$ remain small for an extended period of time, which, in turn, delays the onset of the above instabilities. 
We further expect that the above instabilities are delayed more, if the higher-order terms of the $1/Z$ expansion are included. This subject, however, is beyond the scope of the present work.

\subsection{Ensembles of initial conditions}

\subsubsection{Random spin ensembles}

We define quite generally any one-site reduced matrix density associated with the spin 
direction ${\bf s}_m$ of a quantum state:
\begin{eqnarray}
\hat\rho_m (t=0)= |{\bf s}_m \rangle \langle {\bf s}_m|
=
\frac{\hat 1_m + 2 {\bf s}_m.\hat {\bf S}_m}{2}\,.
\end{eqnarray}
We set the normalization to $|{\bf s}_m|^2=1$ in order to have the density matrix representing a pure state. 
We then define the ensemble of density matrices by adopting the uniform probability distribution  
of all possible orientations of vector  ${\bf s}_m$. This operation is accomplished using the identity:
\begin{eqnarray}
\hat 1_m = 2\oint \frac{d{\bf s}_m}{4\pi} 
\hat \rho_m (t=0) \, ,
\end{eqnarray}
where we define the integration over every spin direction. 
Using these notations, the spin operators along the $x$ axis can be rewritten as:
\begin{eqnarray}
\hat S_m^x =\hat S_m^x(t=0)= 3\oint \frac{d{\bf s}_m}{4\pi} 
\hat \rho_m (t=0) s_m^x \,.
\end{eqnarray}
This definition implies the following initial expectation value for one realization of initial conditions: 
\begin{eqnarray}\label{init}
{\bf S}_m(t=0)&=&\tr_m(\hat \rho_m(t=0){\bf \hat S}_m)
= \,\langle {\bf s}_m|{\bf \hat S}_m|{\bf s}_m \rangle={\bf s}_m/2 \,.
\end{eqnarray}
As a consequence, the autocorrelation function (\ref{correlation})  can be 
rewritten as:
\begin{eqnarray}\label{autavg}
\tr[\hat S^x(t) \hat S^x]
&=& 
2^{L-1} 3 \prod_m \oint \frac{d{\bf s}_m}{4\pi} 
\tr[ \hat S^x(t) \prod_m \hat \rho_m (t=0)] \sum_{m'} s_{m'}^x
\nonumber \\
&=& 
2^{L-1} 3 \prod_m \oint \frac{d{\bf s}_m}{4\pi} 
\sum_{m,m'} S^x_{m}(t) s_{m'}^x \,,
\end{eqnarray}
where $L$ is the number of spins.
For the correlation terms, we use the initial conditions 
\begin{equation}
M_{mn}^{\alpha\beta}(0) =0.
\label{M0}
\end{equation} 

Thus, the functions $S^x_m(t)$ become functionals of the set  $\{{\bf s}_m \}$, which can be obtained 
by solving the system (\ref{S}) and (\ref{M}) with the initial conditions given by Eq.(\ref{init}, \ref{M0}). 
For each set of initial conditions ${\bf s}_m$, 
we determine $S^x_m(t)$ and then carry out the integral over all initial configurations.
In comparison to the classical description,
the inclusion of the correlation terms ensures 
the exact recovery of the short time dynamics, 
more precisely the second order moment term $C(t)=1-\frac{(J^z-J^y)^2}{2}\frac{t^2}{2!}+...$ derived in appendix \ref{A3}.

These considerations can be generalized for a normalization $|{\bf s}_m|$ chosen arbitrarily. 
When $|{\bf s}_m|$ is less than one, the reduced density matrix of a given spin becomes mixed but, 
when $|{\bf s}_m|$ is greater than one, it is difficult to interpret the resulting density matrix 
physically, because one of its eigenvalues is negative, 
while the other is greater than 1. Yet, such states may be used in the context of 
an ensemble average by a straightforward generalization of (\ref{autavg}) with the help of a renormalization.
Such an approach can be justified by noting that, when a classical-spin limit is taken, the simulated system 
is no longer a lattice of spins 1/2 --- rather it can be thought of as a lattice of large quantum spins. 
Therefore, one should concentrate on the average spin polarizations rather than on the density matrices. 
We explore such a renormalization of $|{\bf s}_m|$ as follows.

The ensemble based on the ``natural'' normalization of initial conditions $|{\bf s}_m|=1$ imposes the individual quantum spin 
fluctuations $\langle \delta {\hat {\bf S}}_m^2 \rangle= 3/4 - {\bf S}_m^2=1/2$ associated with a pure state.  We also consider the value 
$|{\bf s}_m|=\sqrt{3}$, which implies no initial quantum fluctuations
$\langle \delta {\hat {\bf S}}_m^2 \rangle= 0$, in a close correspondence with classical simulations. 
The normalization $|{\bf s}_m|=\sqrt{3}$  exactly reproduces  the second moment  already 
at the level of the classical equations as shown in a previous work \cite{Elsayed-15}, thereby leading to a qualitatively 
correct behavior of $C(t)$ for a spin-1/2 chain. It was also shown analytically in Ref.\cite{Lundin-77}, 
that, in the limit of the infinite number of interacting neighbours, correlation functions $C(t)$ computed classically 
and quantum-mechanically become identical --- consequence of the fact that commutators for quantum spins and 
Poisson brackets for classical spins have essentially the same structure\cite{deWijn-13,greg} and, as a result, 
lead to the same expressions for the time derivatives of $C(t)$.

When correlation terms $M^{\beta\gamma}_{mr}$ are included, the 
second moment of $C(t)$ (second derivative at $t=0$)  is always reproduced exactly irrespective of the initial normalization of $|{\bf s}_m|$. 
The idea behind the use of the normalization $|{\bf s}_m|=\sqrt{3}$ in the simulations including correlations 
$M^{\beta\gamma}_{mr}$  is to reproduce the second moment of $C(t)$ already 
without $M^{\beta\gamma}_{mr}$, so that the growth of $M^{\beta\gamma}_{mr}(t)$ is additionally delayed.

\subsubsection{The z-basis ensemble}

In addition to the random ensemble defined in the preceding subsection, we also consider an anisotropic ensemble 
with initial spin polarizations oriented approximately along the $z$-direction. Such an ensemble might be more adequate 
for Hamiltonians commuting with the $z$-projection of the total spin, especially when $J^z_{m,n}$ is the largest coupling constant. 
The initial orientations of ${\bf s}_m$ cannot simply be all along the $z$-direction, 
because, if they were, then each realization of such initial conditions would correspond to an unstable fixed point of the dynamics governed by Eqs.(\ref{S},\ref{M}).  
For this reason, we rotate the z-axis about the y-axis by a vanishing angle $\epsilon \rightarrow 0$, 
so that the x-projections of spins acquire small numerically tractable values:


\begin{eqnarray}
\hat S^x_\epsilon=\frac{\hat S^x-\epsilon \hat S^z}{\sqrt{1+\epsilon^2}} \, ,
\quad \quad
\hat S^z_\epsilon=\frac{\hat S^z+\epsilon \hat S^x}{\sqrt{1+\epsilon^2}}
\end{eqnarray}

We define the ensemble by the initial set of values ${\bf s}_m^{\pm}=\pm (\epsilon, 0 ,1)/\sqrt{1+\epsilon^2}$, which leads to the following set of reduced density matrices:
\begin{eqnarray}
\hat\rho_m^{\pm} (t=0)= |{\bf s}_m^\pm \rangle_m \langle {\bf s}_m^\pm|
=
\frac{\hat 1_m + 2 {\bf s}_m^\pm.\hat {\bf S}_m}{2} \,.
\end{eqnarray}
Using the commutation relation $[\hat S^z, \hat H]=0$, we then express the autocorrelation as:
\begin{eqnarray}
\tr[\hat S^x(t) \hat S^x]
&=&\lim_{\epsilon \rightarrow 0} \tr[\hat S^x(t) \hat S^z_\epsilon]/\epsilon
\nonumber \\
&=& \lim_{\epsilon \rightarrow 0} \sum_{\{{\bf s}_m^\pm\}}
\sum_{m,m'} \tr[ \hat S^x(t) \prod_m \hat \rho_m (t=0)] s_{m'}^{z \pm} /(2\epsilon)
\nonumber \\
&=& \lim_{\epsilon \rightarrow 0} \sum_{\{{\bf s}_m^\pm\}}
\sum_{m,m'} S^x_m(t) s_{m'}^{z \pm} /(2\epsilon) \,.
\end{eqnarray}


\section{Tests for one-dimensional chains}
\label{tests}

In our numerical tests, we restrict ourselves to a periodic spin-1/2 chain with $L=12$ sites. We use the values $J^z=0.82$, $J^x=J^y=-0.41$ 
representing the ratio of the coupling constants between nearest-neighbor spins typical of NMR settings 
\cite{Elsayed-15,VanVleck-48,Lowe-57,Abragam-61}. In such a case, the $z$-projection of the total spin  commutes with the Hamiltonian. 
The reference plots representing fully quantum dynamics are obtained from an exact diagonalization.

Our simulations were subject to the dynamical instabilities mentioned earlier.
We found that some initial conditions lead to stable solutions for longer times, while others create instabilities faster.
In the results presented in Figs. 1-4, a tiny fraction of completely diverging solutions was not included in the averaging.

The test for the random spin ensemble with $|{\bf s}_m|=1$ is presented in Fig. 1. It shows a rather noticeable discrepancy 
with the reference plot. In comparison, the test for the random spin ensemble with  $|{\bf s}_m|=\sqrt{3}$ shown 
in Fig. 2 exhibits a much better initial agreement. 
However, this initial performance was still not better than that of 
purely classical simulations reported in Ref.\cite{Elsayed-15}. At longer times, the individual solutions generated with 
both $|{\bf s}_m|=1$ and   $|{\bf s}_m|=\sqrt{3}$ ensembles begin exhibiting singular  behavior leading to an increasingly 
poor convergence of the statistical averaging procedure.

The test for the $z$-basis ensemble with $|{\bf s}^\pm_m|=1$ is shown in Fig. 3.  Here the averaging was carried out over all 
possible $2^{L}=4048$ initial conditions. This discrete summation procedure appears to be more efficient than that for 
the isotropic random spin ensemble. Yet, for $t>2$ a noticeable discrepancy sets in. We also tested the $z$-basis 
ensemble with $|{\bf s}^\pm_m|=\sqrt{3/2}$, which would lead to the correct second moment, without the contribution 
from the correlation term. The resulting initial performance has improved and, in fact, become arguably better 
than that of the purely classical simulations of Ref.\cite{Elsayed-15}. Yet the agreement with the reference 
plot at longer times remains unsatisfactory.

\begin{figure}
 \begin{center}
 \includegraphics[width=12cm]{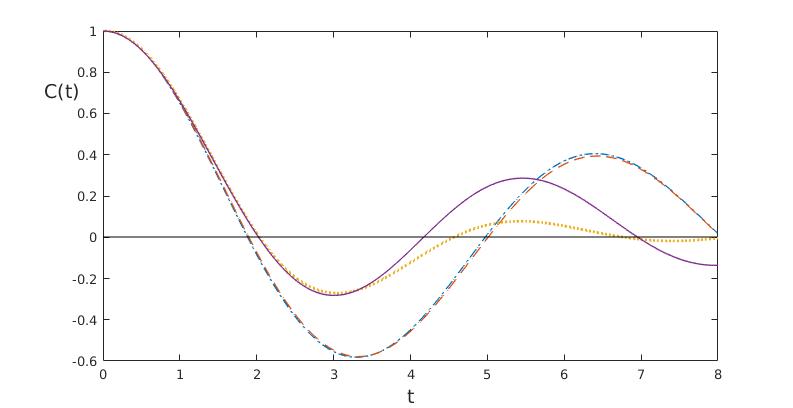}
 \caption{Correlation function (\ref{correlation}) obtained with the random spin ensemble using  normalization $|{\bf s}_m|=1$ for a chain of  12 spins 1/2 described in the text. Red dashed line is obtained for 1000 realizations of initial conditions, blue dot-dashed line for 500 realizations. (The difference between the two lines illustrates the statistical uncertainty.)
 For comparison, solid magenta line represents exact quantum result, and yellow dotted line represents purely classical simulations with  $|{\bf s}_m|=\sqrt{3}$. }
 \end{center}
\end{figure}

\begin{figure}
 \begin{center}
 \includegraphics[width=12cm]{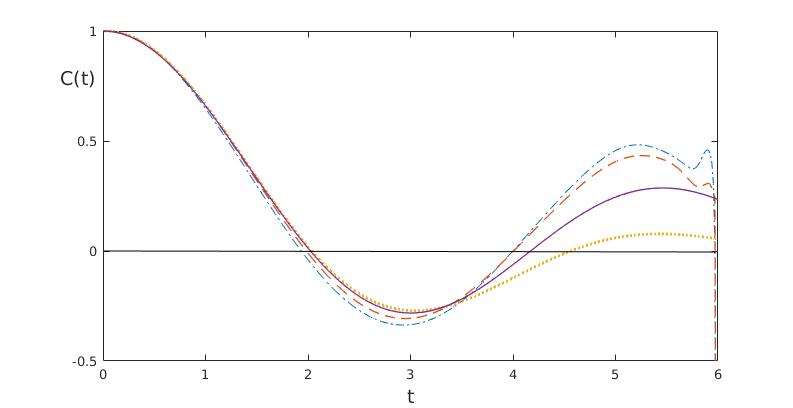}
 \caption{Same as Fig. 1 but with the normalization for the random spin ensemble  
  $|{\bf s}_m|=\sqrt{3}$} and with $3000$ realizations of the initial conditions for the red dashed line and 1500 realizations for the  blue dot-dashed line.
 \end{center}
\end{figure}

\begin{figure}
 \begin{center}
 \includegraphics[width=12cm]{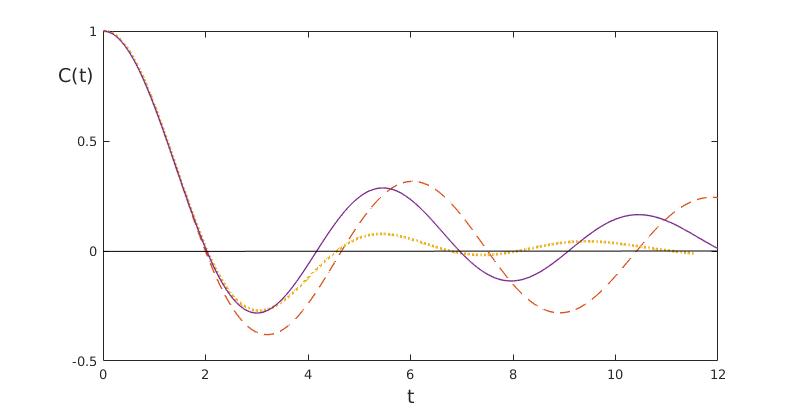}
 \caption{
 Correlation function (\ref{correlation}) obtained with the z-basis ensemble using  normalization $|{\bf s}^\pm_m|=1$ for a chain of  12 spins 1/2 described in the text. Red dashed line is obtained for $2^L=4048$ realizations of initial conditions.
 For comparison, solid magenta line represents exact quantum result, and yellow dotted line represents purely classical simulations with  $|{\bf s}_m|=\sqrt{3}$.}
 \end{center}
\end{figure}

\begin{figure}
 \begin{center}
 \includegraphics[width=12cm]{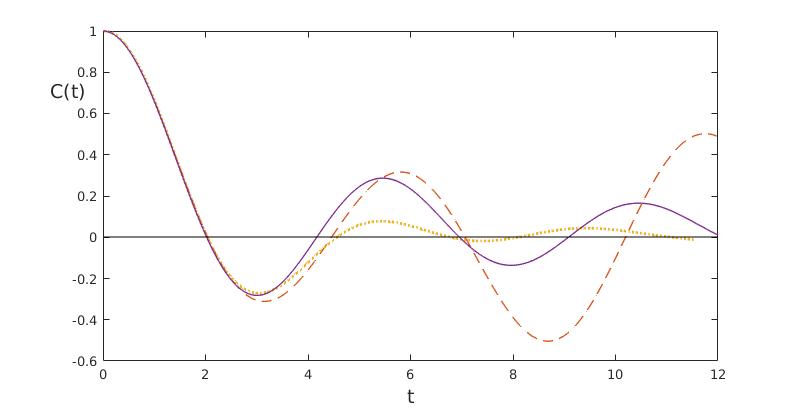}
 \caption{Same as Fig. 3 but with the normalization for the $z$-basis ensemble  $|{\bf s}^\pm_m|=\sqrt{3/2}$.}
 \end{center}
\end{figure}

\section{Summary and outlook}

We analyzed the performance of the large coordination number expansion method with the purpose of modeling 
NMR free induction decays in solids. The method exhibited promising analytical structure. However, despite 
certain improvements at the initial and intermediate times,
the numerical tests indicate that the first-order approximation within this method applied to spin-1/2 chains 
is still not sufficient to outperform  the existing classical calculations. 
A possible reason for this lies in the sensitivity of the method to the choice of the ensemble of initial conditions.  
To conclude, the application of 
the large coordination number expansion  in the NMR context requires further investigations including 
higher-order terms, different ensembles of initial conditions, and, perhaps, different ways of 
including the quantum corrections to the classical equations of motion.

\section*{Acknowledgements}
This work was supported by a grant of the Russian Science Foundation (Project No. 17-12-01587).

\section{Appendix}

\subsection{Two spin cases} \label{A2}

In the case of two spins $m$ and $n$, we set $Z=1$ and 
the system of Eqs.(\ref{S}) and Eqs.(\ref{M}) simplifies into:
\begin{eqnarray}\label{S2}
 \partial_t S_{m}^\alpha
&=& \displaystyle J_{mn}^{\gamma}\epsilon^{\alpha\gamma\beta}
(S_{m}^\beta  S_{n}^\gamma + M^{\beta\gamma}_{mn})   \, ,
\nonumber \\
 \partial_t M_{mn}^{\alpha\beta} 
&=& 
J_{mn}^\gamma \{
\epsilon^{\alpha \gamma\delta }[(\delta^{\beta \delta }/4-S_n^{\beta} S_n^\delta)S_m^\gamma- M_{nm}^{\delta \gamma} S_{n}^\beta]
\nonumber \\
&+&
\epsilon^{\beta \gamma \delta }[(\delta^{\alpha \gamma}/4-S_m^\alpha S_m^\gamma)S_n^\delta  -M_{nm}^{\delta \gamma} S_{m}^\alpha]\}
\,.
\end{eqnarray}
These two set of equations can be combined into
\begin{eqnarray}
 \partial_t (M_{mn}^{\alpha\beta} + S_m^\alpha S_n^\beta)
&=& 
\frac{J_{mn}^\gamma}{4}
\epsilon^{\alpha \gamma\beta} S_m^\gamma
+\frac{J_{mn}^\alpha}{4}
\epsilon^{\beta \alpha \delta }S_n^\delta 
\, .
\end{eqnarray}
The term inside the first time derivative has the same form as the right-hand-side 
of Eq.(\ref{S2}). 
Taking  the second time derivative of (\ref{S2}), we  eliminate  this term to 
obtain for the $x$ component:
\begin{eqnarray}
( \partial_t^2 + \omega_x^2) (S_{m}^x+S_n^x)=0
\end{eqnarray}
where $\omega_x=\frac{|J_{mn}^{y}-J_{mn}^{z}|}{2}$ is frequency of oscillation with the solution 
$S^x(t)=S^x(0)\cos(\omega_x t)$.
Therefore, up to the first order in the large coordination number expansion, we recover the exact
purely oscillatory autocorrelation function 
$C(t)=\cos(\omega_x t)$.

\subsection{Second moment of $C(t)$} \label{A3}

The initial behavior of $C(t)$ can be determined by expanding it 
up to the second order in $t$, thereby obtaining the second moment as a coefficient in front of $t^2/2$:
\begin{eqnarray}\label{exact2nd}
\frac{\tr\left(\hat S^x(t) \hat S^x\right)}
{\tr\left(\hat S^{x\,2}\right)}
&=&1+\frac{\tr\left([\hat H,\hat S^x]^2\right)}
{\tr\left(\hat S^{x\,2} \right)}\frac{t^2}{2!}+ \dots
\simeq 
1-
\frac{(J_z-J_y)^2}{2}
\frac{t^2}{2!}
\,.
\end{eqnarray}
This exact result can be compared with the one obtained 
from our approximations. For this purpose, we solve 
Eqs.(\ref{S}) through a perturbation expansion 
$S^\alpha_m(t)={S^\alpha_m}+{S^\alpha_m}^{(1)}t+{S^\alpha_m}^{(2)}t^2/2!+\dots$ and 
$M^{\alpha\beta}_{mn}(t)={M^{\alpha\beta}_{mn}}^{(1)}t+\dots$ and for the autocorrelation function:
\begin{eqnarray}\label{2nd}
\tr[\hat S^x(t) \hat S^x]^{(2)}
&=& 
2^{L-1} 3 \prod_m \oint \frac{d{\bf s}_m}{4\pi} 
\sum_{m,m'} {S^x_{m}}^{(2)} s_{m'}^x \, .
\end{eqnarray}
After 
the expansion, we obtain the relevant set of equations:
\begin{eqnarray}
{S_{m}^y}^{(1)}
&=& 
\sum_{u=\pm 1} \left(-J^{x} 
S_{m}^y  S_{m + u}^x 
+ J^z S_{m}^x  S_{m + u}^z \right)\, ,
\\
{S_{m}^z}^{(1)}
&=& 
\sum_{u=\pm 1} \left(-J^{y} 
S_{m}^x  S_{m + u}^y 
+ J^x S_{m}^y  S_{m + u}^x \right)
\\
{S_{m}^x}^{(2)}
&=& 
\sum_{u=\pm 1} \biggl[ -J^{z}
(S_{m}^{y(1)} S_{m +u}^z + 
S_{m}^{y} S_{m +u }^{z(1)}+M_{m, m +u}^{yz(1)}) \, ,
\nonumber \\
&+&J^{y}(
S_{m}^{z(1)}  S_{m +u }^y +
S_{m}^z  S_{m +u }^{y(1)}+M_{m, m+u}^{zy(1)})\biggr]
\\
M^{yz(1)}_{m, m \pm 1}
&=&-J^x S_m^x S_m^y S_{m\pm 1}^y -J^y(1/4-{S_m^y}^2)S^x_{m \pm 1}
\nonumber \\
&+&J^z(1/4-{S^z_{m \pm 1}}^2)S^x_m + J^x S^x_{m \pm 1} S^z_{m\pm1}S^z_m \, .
\end{eqnarray}
Solving this set and inserting the result into (\ref{2nd}), we recover, after setting 
$S^i_m=s^i_m/2$, the 
exact result (\ref{exact2nd}).

On the other hand, if we neglect the 
quantum pair correlation terms $M^{yz(1)}_{m, m \pm 1}$ for the random spin ensemble, we obtain 
for any normalization $|{\bf s}_m|$:
\begin{eqnarray}\label{clas2nd}
\frac{\tr\left(\hat S^x(t) \hat S^x\right)}
{\tr\left(\hat S^{x\,2}\right)}
&\approx&
1-
\frac{(J^z-J^y)^2|{\bf s}_m|^2}{6}
\frac{t^2}{2!} \, .
\end{eqnarray}
The random spin ensemble with the normalization $|{\bf{s}}_m|=\sqrt{3}$  
leads to the correct second moment (\ref{exact2nd}) without the quantum terms $M^{\alpha \beta}_{m n}$ but differs from the ``natural'' normalization $|{\bf{s}}_m|=1$ 
by a factor $\sqrt{3}$.
In contrast, for the $z$-basis ensemble, we obtain instead 
\begin{eqnarray}\label{clasz}
\frac{\tr\left(\hat S^x(t) \hat S^x\right)}
{\tr\left(\hat S^{x\,2}\right)}
&\approx&
1-
\frac{J^z(J^z-J^y)|{\bf s}_m^\pm|^2}{2}
\frac{t^2}{2!} \, .
\end{eqnarray}
Since the coupling values are chosen such that $J^z-J^y=3J^z/2$, 
the normalization is fixed to 
$|{{\bf s}}_m^\pm|=\sqrt{3/2}$ in order to reproduce the exact second moment.

%

%




\end{document}